\documentclass[aps, prl,
 amsmath,
 amssymb,
 reprint,
showpacs,
showkeys,
]{revtex4-1}
\usepackage{graphicx}
\pdfoutput=1
\usepackage{dcolumn}
\usepackage{bm}

\begin{document}
\title{Nonlinear dynamics of self-sustained supersonic reaction waves:
Fickett's detonation analogue}
\author{M. I. Radulescu}
   \email{matei@uottawa.ca}
\author{J. Tang}
\affiliation{Department of Mechanical Engineering, University of Ottawa}
\date{\today}

\begin{abstract}
The present study investigates the spatio-temporal variability in the dynamics of self-sustained supersonic reaction waves propagating through an excitable medium.  The model is an extension of Fickett's detonation model with a state dependent energy addition term.  Stable and pulsating supersonic waves are predicted. With increasing sensitivity of the reaction rate, the reaction wave transits from steady propagation to stable limit cycles and eventually to chaos through the classical Feigenbaum route. The physical pulsation mechanism is explained by the coherence between internal wave motion and energy release. The results obtained clarify the physical origin of detonation wave instability in chemical detonations previously observed experimentally.  
\end{abstract}
\maketitle

Self-sustained waves propagating through excitable media can give rise to complex patterns (e.g., spirals, cells, target shapes) and generally show large spatio-temporal variability in their dynamics.  While subsonic wave propagation is usually modelled by reaction-diffusion coupling, supersonic self-sustained waves rely on the coupling between energy release in the media and the resulting mechanical waves (elastic, compression waves, etc...). The latter class of waves will henceforth be called \textit{detonations}, in analogy to the self-sustained supersonic waves commonly encountered in reactive gases, combustion of dust particles in air and condensed phase energetic materials \cite{Fickett&Davis1979}. Such detonations can also appear in media sustaining thermo-nuclear reactions, and have been hypothesized to be the decomposition mode of white dwarfs undergoing supernova explosions \cite{Arnett1969}. Such supersonic self-sustained waves have also been observed in a wide variety of elastic excitable media, such as Burridge-Knopoff models of frictional sliding, electronic transmission lines, and active optical waveguides \cite{Cartwrightetal1997}. Likewise, phase change waves \cite{Gulenetal1994}, traffic jams \cite{Flynnetal2009} and shallow water waves \cite{Kasimov2008} also share the same characteristics of detonations, whereby the arrival of mechanical waves induces a change in the material state, conducive to the release of energy, which in turn modifies, or sustains the wave motion.

A simple and elegant model for self-sustained supersonic reaction waves has been introduced by Fickett \cite{Fickett1979, Fickett1985}. This model is known to reproduce qualitatively many dynamic traits of chemical detonations, such as the structure of the self-sustained wave, initiation transients and response to boundary losses (see \cite{Fickett1985}). Its mathematical simplicity offers a much simpler framework to study detonations than the reactive Euler equations, as is the case for chemical detonations. Its simplicity also permits to consider this model as a unifying model to the wide variety of waves in excitable media mentioned above. The present study is an investigation of its spatio-temporal non-linear dynamics, which have not been addressed in previous studies.

Indeed, in gaseous chemical detonations, experiments have demonstrated the large spatio-temporal variability in the wave propagation.  In multiple dimensions, multi-scale cellular patterns have been observed \cite{Fickett&Davis1979}, while in a single space dimension, a pulsating instability has been observed \cite{Edwards&Morgan1977}. Numerical simulations of the reactive Euler equations used to model one-dimensional pulsating detonations have shown the universality in the wave dynamics \cite{Ngetal2005}. As the sensitivity of the reaction rates is increased, stable travelling waves become oscillatory, and subsequently develop a hierarchy of period doubling bifurcations appearing according to Feigenbaum's scaling \cite{Feigenbaum1983}, and eventually become chaotic. At present, because of the complexity of the reactive Euler equations and resulting dynamics, neither the mechanism of the one-dimensional pulsating instability nor the reason for the universality in the period-doubling detonation dynamics are understood. In this paper we study the wave stability predicted by Fickett's model whose simplicity allows for analytical investigation of period-doubling and chaotic dynamics.

Fickett's mathematical model for detonations is an extension of the inviscid Burgers' equation to the reactive case, that is:
\begin{align}
\partial_t \rho + \partial_x p=0 \label{eq:Fickett} \\
\partial_t \lambda_r =r\left( \rho, \lambda_r \right) \label{eq:rate}
\end{align}
where all fields are defined on $(x,t)$. The model is formulated in Lagrangian coordinates, where $x$ denotes a material coordinate, while $t$ represents time \cite{Fickett1985}.  The variable $\rho$ has the meaning of density in the reactive analogue. The flux term $p$ in \eqref{eq:Fickett} has the meaning of pressure, see  \cite{Fickett1985}. For simplicity, we choose the generic equation of state proposed by Fickett, that is
\begin{align}
p=\frac{1}{2} \left(\rho^2 + \lambda_r Q \right) \label{eq:pressure}
\end{align}
The available energy in the excitable medium is $Q$ and $\lambda_r$ denotes the progress variable of the energy process, ranging from 0 (un-reacted) to 1 (reacted).  Equation \eqref{eq:rate} describes the evolution of the energy release progress for each particle with Lagrangian coordinate $x$; a model for this reaction term will be described below. Note that by setting $Q$ to zero, one recovers the well-studied inviscid Burgers' equation \cite{Whitham1974}.

More insight into the interplay between wave motion and energy addition can be obtained by recognizing that the system of equations \eqref{eq:Fickett} and \eqref{eq:rate} is hyperbolic.  It can be shown that the characteristic form can be written as:
\begin{align}
\frac{d p}{d t}=rQ \; \mathrm{\quad along \quad}\;  \frac{dx}{dt}=\rho \label{eq:charplus}\\
\frac{d \lambda_r}{d t}=r \; \mathrm{\quad along \quad} \; \frac{dx}{dt}=0\label{eq:char0}
\end{align}
From \eqref{eq:charplus}, we deduce that the system exhibits waves propagating forward with speed $dx/dt=\rho$. The wave communicates changes in pressure amplitude only in the positive $x$ direction.  The amplitude of the wave varies with the heat addition $Q$ at the rate $r$. Hence the model describes the physical property that waves may amplify in the presence of energy release according to Rayleigh's criterion, i.e., if the energy release is in phase with the wave motion. The second family of characteristics in \eqref{eq:char0} gives the rate of energy release along a particle path. The physical picture emerging is thus the reactivity set out along particle paths modifies the strength of waves propagating forward. Through the coupling of the reaction rate (which we will ascribe below) to wave strengths, the feedback loop is closed.  Note that contrary to the reactive Euler equations used to model chemical detonations in fluids, which admit three sets of waves \cite{Leung2010}, the analogue predicts two, as rear facing pressure waves are absent.  This fundamental simplification permits to analyse the detonation problem in an analytically tractable system of equations, unlike the reactive Euler equations.

The system admits a coherent self-propagating travelling wave solution having the properties of a detonation \cite{Fickett1985}. Although the details are available \cite{Fickett1985}, we briefly describe its steady solution, as it serves as our starting point for the stability analysis.  We seek a travelling wave solution to the system given by \eqref{eq:Fickett} and \eqref{eq:rate}.  The speed of the wave $D$ can be found in terms of the unreacted state $(\rho_0, \lambda_{r0})$ in front of the wave and the reacted state $(\rho_2, \lambda_{r2})$ behind the wave.  For simplicity, and without any loss of generality, we set $\rho_0=0$, $\lambda_{r0}=0$ and $\lambda_{r2}=1$ to model an irreversible exothermic reaction.  We also let $\rho_2$ variable (i.e. the piston problem, see Fickett \& Davis\cite{Fickett&Davis1979}).  Adopting the notation $[\zeta]=\zeta_2-\zeta_0$, the resulting wave speed can be found (see \cite{Whitham1974}) from $D=[p]/[\rho]=\left( {\rho_2}^2+Q \right) /\left( 2 \rho_2 \right)$.

The self-sustained travelling wave solution corresponds to the case where the forward propagating characteristic trailing the wave cannot penetrate the wave structure, and essentially represents an event-horizon.  The speed of this so-called limiting characteristic thus needs to be equal to the detonation speed.  Denoting this special case as the Chapman-Jouguet case (by analogy to the terminology used in chemical detonations \cite{Fickett&Davis1979}) with subscript $CJ$, we require that $\rho_2=D=D_{CJ}$, from which we obtain the CJ speed of the detonation.
\begin{align}
D_{CJ}= \sqrt{Q} \label{eq:Dcj}
\end{align}
Since we are dealing with an inviscid system and the medium develops shocks according to Burgers' equation, the detonation can be assumed to be lead by an inert shock, across which there is no energy release and the density and pressure change discontinuously. We will denote the state behind the shock with a subscript 1 (known as the von Neumann state in the chemical detonations).  For a non-reactive shock satisfying the weak form of the inert inviscid Burgers equation, we obtain $\rho_1=2D$.

The structure of the detonation wave across which energy is deposited at a finite rate is obtained by integrating the governing equations.  The steady wave solution can be obtained by first introducing the change of coordinates $(\zeta=x-D_{CJ}t-x_0, t'=t)$ which defines a local coordinate system moving with the steady detonation. Making the formal change of variables and setting the time derivatives equal to zero in order to obtain the steady solution, we obtain:
\begin{align}
\frac{d}{d\zeta} \left(\frac{1}{2} \rho^2 -D_{CJ}\rho+ \frac{1}{2}\lambda_r Q \right)=0 \label{eq:Steady1}\\
\frac{d}{d\zeta} \left(D_{CJ}\lambda_r \right) = r  \label{eq:Steady2}
\end{align}
This system is integrated from the shock, with the inert shock state $\rho=\rho_1$ and $\lambda_r=0$ as boundary condition at $\zeta=0$, once the rate $r(\rho,\lambda_r)$ is given.

In the present work, we propose and investigate a reaction model that is sufficiently simple to permit us to explain the unsteady period doubling wave dynamics of detonations and sufficiently rich to capture the main non-linearity in travelling waves in excitable media, which is the coupling between (shock) wave motion and exothermicity induced by the shock. We thus extend the models introduced in \cite{Fickett1985} and \cite{Hall&Ludford1987}, for which instability was predicted from linear stability analysis, to a simple generic two step model.  Following the leading shock, we assume the existence of a thermally neutral induction delay, whose duration depends on the strength of the shock.  This is an excellent assumption for activated chemical reactions \cite{Fickett&Davis1979}, but can represent the excitability of any medium. Following the induction process, we assume an exothermic reaction that proceeds at a state-independent constant rate.  The latter choice was selected in order to decouple the activation of the reactions with the duration of the exothermic stage, in order to clearly isolate the physical phenomena governing wave motion and the instability mechanism and avoid the singularity in \cite{Fickett1985} associated with square wave detonations. The resulting generic induction-reaction model we propose is thus:
\begin{align}
\partial_t \lambda_i = -K_i H(\lambda_i) e^{\alpha \left( \frac{\rho}{2D_{CJ}} - 1 \right)} \label{eq:rate1}\\
\partial_t \lambda_r =  K_r \left(1-H(\lambda_i)\right) \left(1-\lambda_r \right)^{\nu}  \label{eq:rate2}
\end{align}
where $K_i$ and $K_r$ are constants controlling the times scales of the induction and reaction zones, respectively. The Heaviside function $H(\cdot)$ controls the timing of the onset of the second exothermic reaction, which starts when the induction variable $\lambda_i$ reaches 0.  Ahead of the shock, $\lambda_i=1$ and  $\lambda_r=0$.  We are also assuming that the reactions are only activated by the passage of the inert leading shock.  The system to be solved is thus \eqref{eq:Fickett}, \eqref{eq:rate1} and \eqref{eq:rate2}.

The reaction model allows for direct analytical derivation of the steady travelling wave solution.  Ahead of the wave in the quiescent zone, we have, $\zeta>0$, $\rho=0$, $\lambda_i=1$ and $\lambda_r=0$.  The induction zone terminates at $\zeta_i=-D_{CJ}/K_i$.  In the induction zone we have $\zeta_i <\zeta < 0$, $\rho=\rho_1=2D_{CJ}$, $\lambda_i=1+{K_i}/({D_{CJ}}\zeta)$, and $\lambda_r=0$. For a reaction order $\nu < 1$ the reaction layer terminates at a finite distance from the shock given by $\zeta_r=\zeta_i-D_{CJ}/(K_r(1-\nu))$.  In the reaction layer, we have, $
\rho=D_{CJ}\left(1+\left(1+(1-\nu )K_r/D_{CJ}(\zeta -\zeta_i)\right)^{\frac{1}{2(1-\nu )}}\right)$ and $\lambda_r=1-\left(1+(1-\nu )K_r/D_{CJ}(\zeta -\zeta_i)\right)^{\frac{1}{1-\nu }}$. 

The non-linear stability of the travelling wave solution of \eqref{eq:Fickett}, \eqref{eq:rate1} and \eqref{eq:rate2} was investigated by numerical integration starting with the steady travelling wave structure as initial condition. The numerical integration uses the fractional steps method, whereby the hydrodynamic evolution and reactive step can be performed separately.  The hydrodynamic step uses an exact first order Riemann solver. Owing to the simplicity of the reactive model, the reactive part of the governing equations is solved in closed form at each time step.  A grid resolution of 256 grid points per detonation wave thickness was used, which ensured the stability boundary was grid independent with an accuracy of $\pm 0.1$ in the value of $\alpha$.
\begin{figure}
\begin{center}
 \includegraphics[width=1\columnwidth]{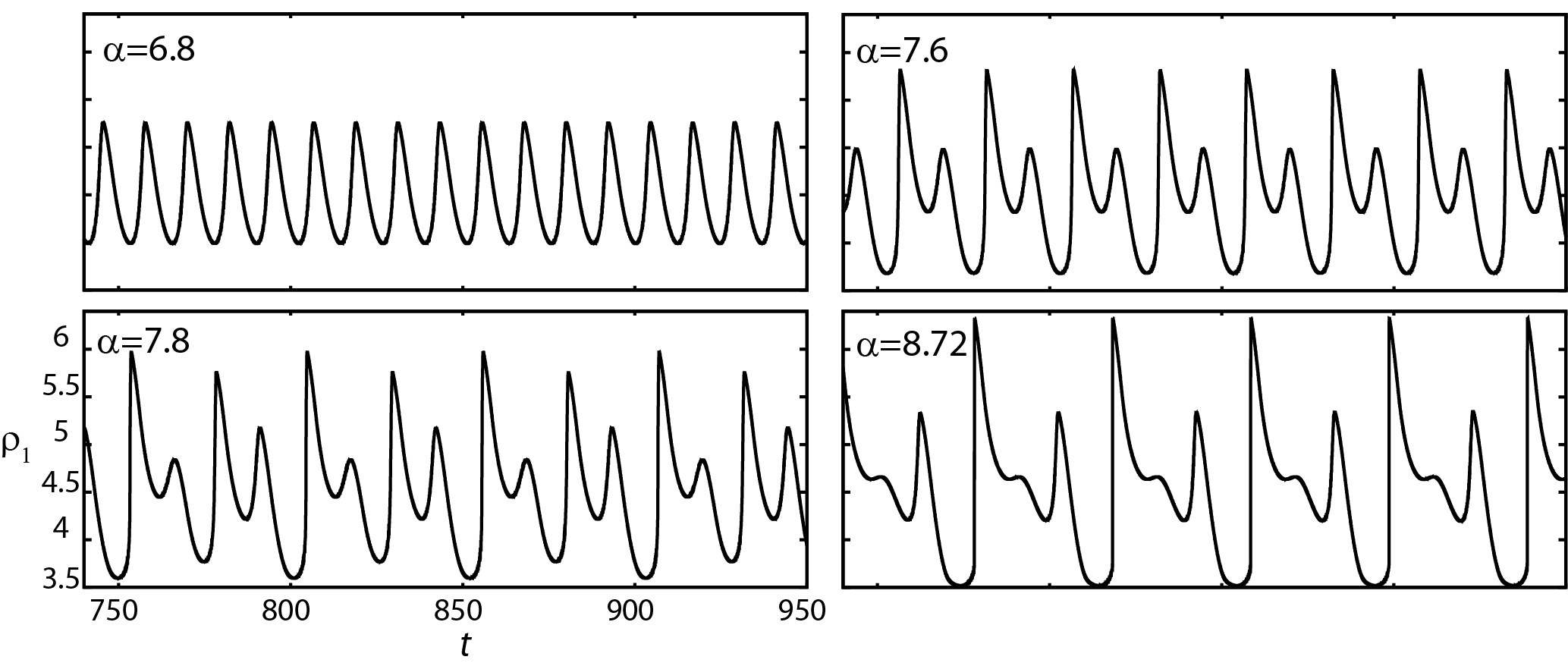}
\caption{Shock amplitude evolution for different values of $\alpha$.} 
\label{fig:fig1}
\end{center}
\end{figure}
The results presented are for parameters, $Q=5$, $K_i=1$, $K_r=2$ and $\nu=0.5$.  Below the critical value $\alpha=5.7$, the steady solution is stable, and propagated with the steady wave structure given above at its constant CJ speed given by \eqref{eq:Dcj}.  Above this critical value the travelling wave solution is unstable and develops a stable limit-cycle, as shown in Fig. \ref{fig:fig1}. As $\alpha$ increases, the amplitude of the pulsations also increases until a period doubling bifurcation occurs at $\alpha = 6.9$.  Fig. \ref{fig:bifurcation} shows the bifurcation diagram. Further increases in $\alpha$ yields another bifurcation at $\alpha = 7.7$, followed by subsequent bifurcations occurring with smaller changes in $\alpha$.  Eventually, the pulsations become chaotic, with isolated values of $\alpha$ for which the dynamics have an odd period; one example is shown in Fig. \ref{fig:fig1}.  This implies the onset of chaos \cite{Ngetal2005}.  The sequence of period doubling bifurcations and onset of chaos is very similar to the non-linear dynamics of chemical detonations \cite{Ngetal2005} and many other non-linear systems.  The results thus clearly highlight that our simple detonation model captures this universality observed in non-linear dynamics of complex systems.
\begin{figure}
\begin{center}
\includegraphics[width=0.9\columnwidth]{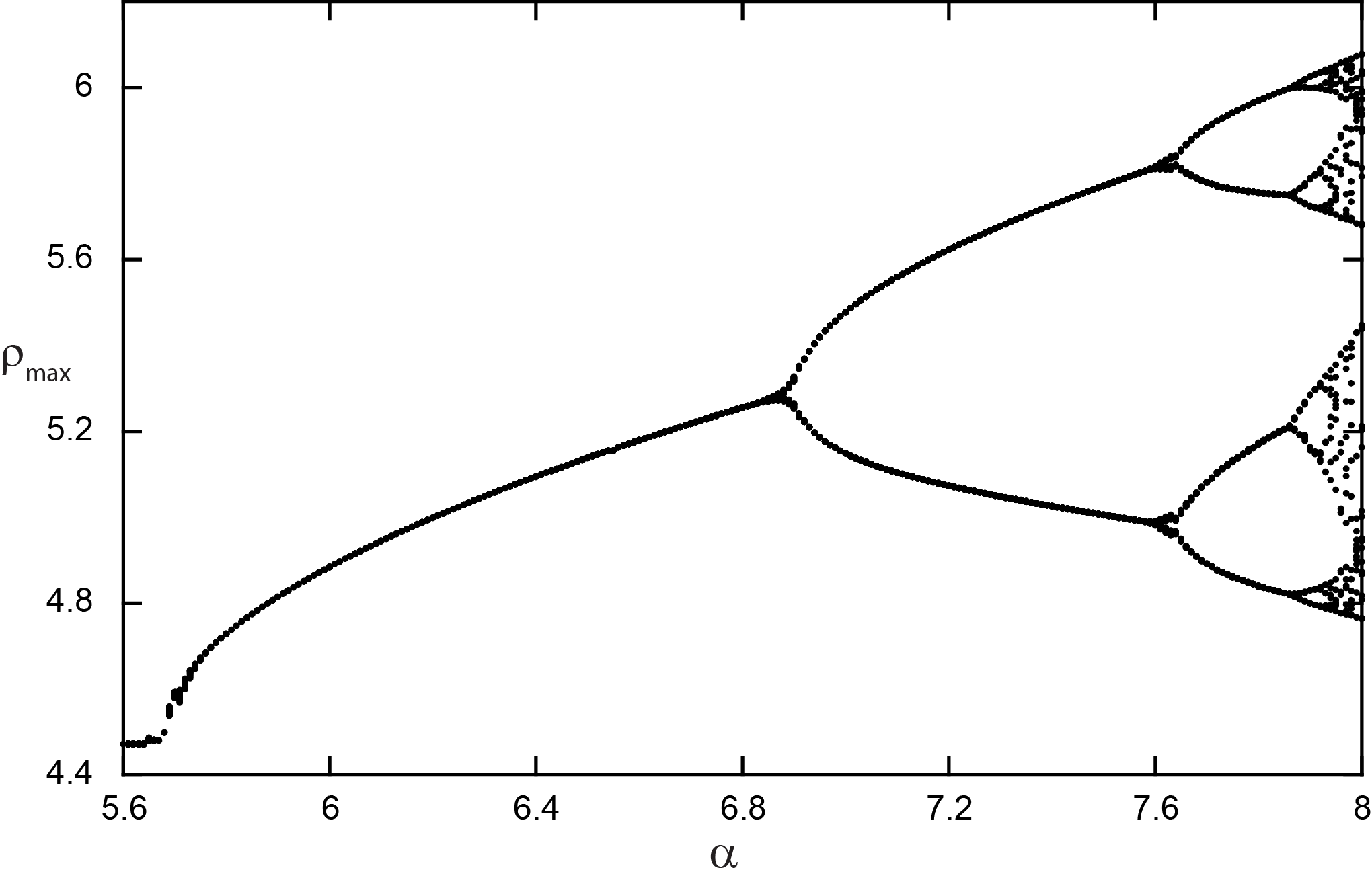}
\caption{Bifurcation diagram of the local maxima of detonation wave strength $\rho_{max}$ for different induction zone sensitivities $\alpha$.} 
\label{fig:bifurcation}
\end{center}
\end{figure}

\begin{figure}
\begin{center}
\includegraphics[width=0.85\columnwidth]{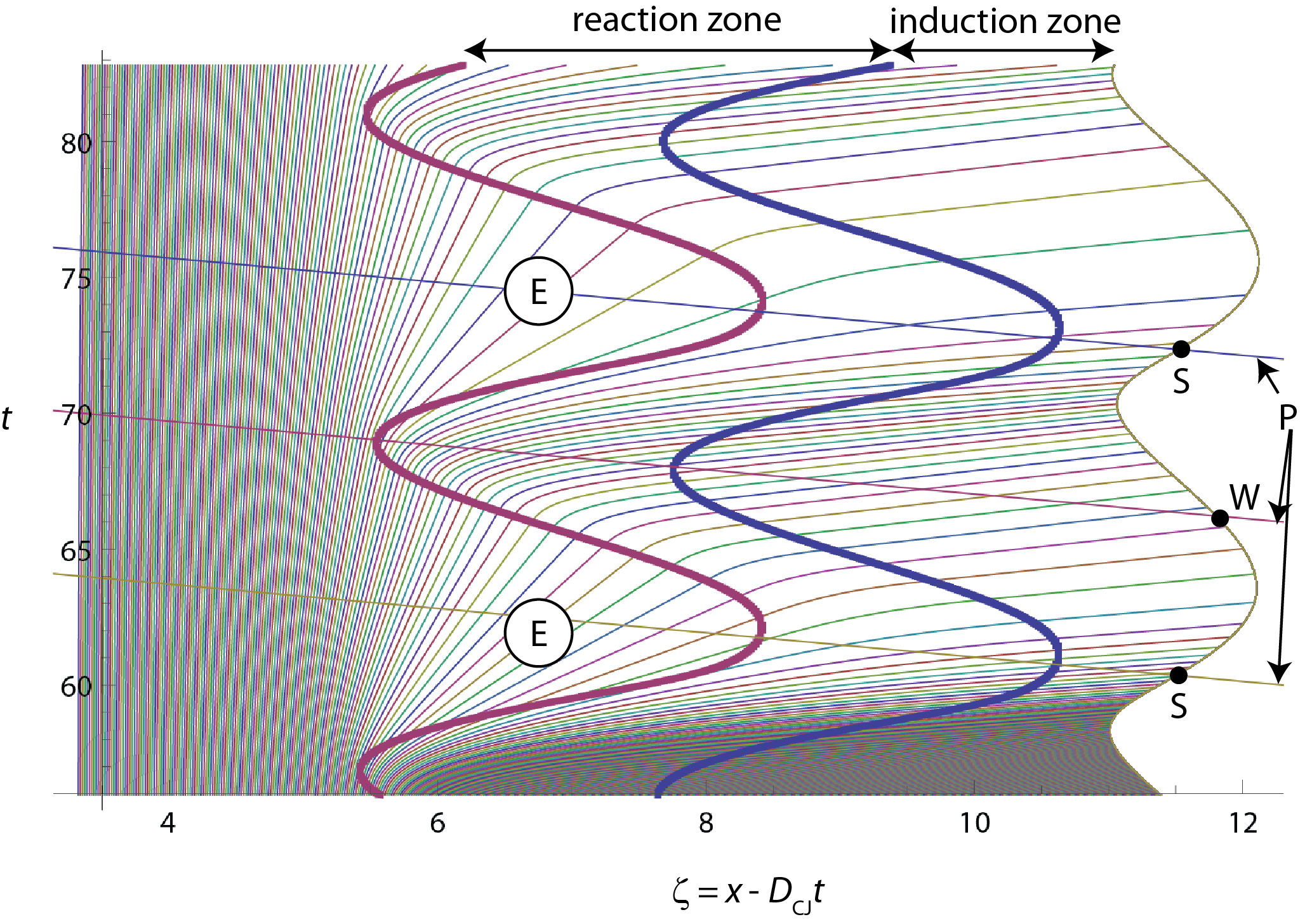}
\caption{Space time diagram illustrating the pressure waves in the reaction zone of a pulsating detonation.} 
\label{fig:fig3}
\end{center}
\end{figure}

The analogue model also allows for a straightforward interpretation of the instability mechanism in detonations. In order to study the non-linear instability mechanism of the proposed detonation analogue, we focused our attention on the single mode limit-cycle obtained for $\alpha=6.8$.  Figure \ref{fig:fig3} illustrates the evolution of the wave structure over approximately two oscillation periods, in the frame of reference of the steady travelling wave. To visualize the dynamics, we reconstructed an (arbitrary) discrete set of pressure waves by integrating \eqref{eq:charplus} starting from arbitrary locations.  We used a predictor-corrector method and interpolated on the solution obtained above. The lead shock front of the detonation corresponds to the locus where these characteristics coalesce.  Behind the oscillating lead shock are the two zones of induction and reaction. Note that the finite (numerical) dissipation at the discontinuity at the rear of the reaction zone makes the characteristics bend somewhat towards the reaction zone in a very narrow region.  Away from this region, by virtue of the characteristic equation \eqref{eq:charplus}, the pressure waves have constant amplitude and speed everywhere except in the reaction zone, where they accelerate owing to the heat release.  By investigation of the characteristic diagram of Fig. \ref{fig:fig3}, the detonation wave structure can be easily interpreted as the coherent wave structure formed by the amplification of forward travelling waves.  These are amplified across the reaction zone and eventually reach the hydrodynamic shock.  Since the onset of the reactions is controlled by the lead shock, the pressure waves continuously see the same reacting field and the self-sustained detonation phenomenon occurs.

The instability mechanism itself can be inferred from the characteristic diagram shown in Figure \ref{fig:fig3}.  First, one can note that the oscillations of the leading shock provide a modulation in the duration of the induction zone and location of the reaction zone.  This is due to the induction time sensitivity on shock state given in \eqref{eq:rate1}. As can be verified in Fig. \ref{fig:fig3}, the points when the shock is strongest (S) correspond to the shortest induction times, which is communicated along the particle paths (P).  Likewise, the weakest shocks (W) yield the longest induction times.  Due to this modulation in the onset of exothermicity, the acceleration and deceleration phases of the lead shock pulsations can be deduced.  The amplification stage of the lead shock corresponds to the arrival of pressure waves that travelled in phase with the energy release zone, which can be identified as the regions where the pressure waves travel almost parallel to the reaction zone band.  Likewise, the deceleration phases correspond to the arrival at the lead shock of waves travelling out of phase with the energy release zone. The waves travelling in phase with the energy release amplify more, and communicate an acceleration to the lead shock. Since this occurs during the lead shock amplification stage, the feedback accentuates the amplification. Likewise, a decelerating shock provides a non-coherent interaction between the forward pressure waves and exothermicity, further promoting the deceleration. In our system, the coherent amplification can be obtained by integrating \eqref{eq:charplus} for a constant rate, which shows that the amplification of a pressure wave is proportional with the residence time of the wave in the energy release zone.  When a pressure wave stays in phase with an energy release zone for longer times, it acquires the most amplification. The pulsation mechanism, which controls the sequence of acceleration and deceleration phases can also be seen in Fig. \ref{fig:fig3}. Following an acceleration stage, the forward characteristics emanating from the rear of the reaction zone, which only \textit{clip} the reaction zone and obtain very little amplification form the expansion waves (E). The expansion waves immediately following the compression waves provide the restoring mechanism for the instability. Note that the same mechanism has been suggested to be at play in chemical detonations \cite{McVey&Toong1971, Leung2010}, although the complexity of the governing equations did not permit to clearly isolate these effects. The much simpler detonation model suggested in the present study provides very similar dynamics and permits a much more accurate physical investigation of the physical mechanisms controlling the instability, namely the coherent amplification of forward facing waves modulated by the onset of reactions.  This is essentially Rayleigh's criterion for (acoustic) wave amplification by coherent energy release.

\bibliography{references}
\end{document}